\documentstyle[preprint,aps]{revtex}
\begin{document}
\draft
%
%


\title {\Large \bf
Phase Separation of Bose-Einstein Condensates}
\author{E. Timmermans \\
Institute for Atomic and Molecular Physics \\
Harvard-Smithsonian Center for Astrophysics \\
60 Garden Street \\
Cambridge, MA 02138}
\date{\today}
\maketitle
\begin{abstract}

       The zero-temperature system of two
dilute overlapping Bose-Einstein condensates is unstable against long 
wavelength excitations if the interaction strength
between the distinguishable bosons exceeds the
geometric mean of the like-boson interaction strengths.  
If the condensates attract each other, the instability is 
similar to the instability
of the negative scattering length condensates.  If the condensates repel, they
separate spatially into condensates of equal pressure.  
We estimate the boundary size, surface
tension and energy of the phase separated condensate system and we
discuss the implications for double condensates in atomic traps.

\end{abstract}



	 As dilute gases, the atomic trap Bose-Einstein condensates
\cite{first} occupy a unique position among the superfluid systems.
One intriguing consequence of their dilute gas nature is the prospect
of studying condensate mixtures, which are the first experimentally realizable
bosonic superfluid mixtures \cite{rem}-\cite{fet}
(the only
other mixture of superfluids is the fermion-boson $^{3}$He--$^{4}$He system). 
Understandably, this prospect has attracted interest
\cite{Ho}-\cite{Walls}, and recently the first observation
of overlapping condensates was reported \cite{overlap}.

	In this paper, we study if and when zero temperature dilute condensates
overlap.  We find that the
homogeneous overlapping condensate
system is unstable against long wavelength excitations if
the strength of the interaction between
the distinguishable bosons exceeds the geometric mean of the like-boson
interaction strengths. In that case, two repelling condensates
spatially separate into single condensates of equal pressure.
The condensates still partially overlap in the boundary region that separates
them.  We estimate the size of the boundary region, 
as well as the corresponding surface tension. 
We show how the homogenous treatment may
be generalized to describe phase-separated large double
condensate systems in a trap.

	The instability of the overlapping condensates manifests itself in the
energy dispersion of the elementary excitations \cite{Walls}, as we show below.
The wavefunctions, $\phi_{1}$ and $\phi_{2}$, of two interacting
condensates satisfy coupled
Gross-Pitaevski equations:
\begin{eqnarray}
i \hbar \dot{\phi}_{1} &=& \left[ - \frac{ \hbar^{2}  \nabla ^{2}}{2 m_{1}}
 - \mu_{1} + \lambda_{1} | \phi_{1} |^{2} \right] \phi_{1} + \lambda
|\phi_{2}|^{2} \phi_{1}
\nonumber \\
i \hbar \dot{\phi}_{2} &=& \left[ - \frac{ \hbar^{2}  \nabla ^{2}}{2 m_{2}}
 - \mu_{2} + \lambda_{2} | \phi_{2} |^{2} \right] \phi_{2} + \lambda
|\phi_{1}|^{2} \phi_{2}  \; \; ,
\label{e:cgpe}
\end{eqnarray}
where $\mu_{j}$ ($j=1,2$) represents the chemical potential of the $j$-bosons.
The interaction
strength values, $\lambda_{j}$ and $\lambda$, are determined by the scattering
lengths for binary collisions of distinguishable bosons: $\lambda_{j} = 4 \pi
\hbar^{2} a_{j} / m_{j}$ and $\lambda = 2 \pi \hbar^{2} a / m_{red}$, where
$m_{red}^{-1} =
m_{1}^{-1} + m_{2}^{-1}$.
The excitations of the static homogeneous condensates,
$\phi_{j}({\bf r},t) = \phi_{j}^{(0)}$, are described by 
fluctuations of the fields,
$\phi_{j}({\bf r},t) = \phi_{j}^{(0)} + \delta \phi_{j}$, which
evolve according to the Gross-Pitaevski
equations (\ref{e:cgpe}), linearized in $\delta \phi$ and 
$\delta \phi^{\ast}$ \cite{remark}.
Decomposing the field fluctuations
into Fourier components,
$\delta \phi_{j} = \sum_{\bf k} c_{j,{\bf k}} \exp(i {\bf k} \cdot {\bf r})$,
we obtain the equations of motion for the c-amplitudes,
\begin{eqnarray}
i \hbar \dot{c}_{1,{\bf k}} &=& \left[k^{2} / 2m_{1} + \lambda_{1} n_{1} \right]
c_{1,{\bf k}} + \lambda_{1} \phi_{1}^{(0)} c_{1,-{\bf k}}^{\ast}
\nonumber \\
&& \; \; \; \; \;
+ \lambda \phi_{1}^{(0)} \left[ \phi_{2}^{(0) \ast} c_{2,{\bf k}}
+ \phi_{2}^{(0)} c_{2, - {\bf k}}^{\ast} \right] ,
\label{e:c}
\end{eqnarray}
where $n_{j} = |\phi_{j}^{(0)}|^{2}$
and where we have used that $\dot{\phi}_{j}^{0} = 0$.
A second equation for $i \hbar \dot{c}_{2}$ is obtained 
by exchanging the $1$ and $2$ subscripts. 
Alternatively, we can introduce
the phase and density of the condensate field, $\phi = \sqrt{\rho}
\exp(i \theta)$, the fluctuations of which, $\rho = n + \delta \rho$ and
$\theta = \theta^{(0)} + \delta \theta$, account for the field fluctuations,
$\delta \phi = \phi^{(0)} \left[ \delta \rho /2n + i \delta \theta \right]$.
The last term of Eq.(\ref{e:c}) then represents a density fluctuation,
$\delta \rho_{\bf k} = \left[ \phi^{(0) \ast} c_{\bf k} + \phi^{(0)}
c_{-{\bf k}}^{\ast} \right] $ whereas the difference,
 $\delta \Pi_{\bf k} =
\left[ \phi^{(0) \ast} c_{\bf k} - \phi^{(0)} c_{-{\bf k}}^{\ast} \right] /2i$,
represents a phase fluctuation.
Multiplying Eq. (\ref{e:c}) by 
$\phi_{1}^{(0) \ast}$ and adding and subtracting the resulting equalities
with the complex conjugate equation, the $\delta \Pi$ and 
$\delta \rho \;$-equations of motion follow: 
\begin{eqnarray}
\hbar \delta \dot{\rho}_{1,{\bf k}} &=& 2 \left[ \hbar^{2} k^{2}/2m_{1} \right]
\delta \Pi_{1,{\bf k}} \; \; \; \; \; \; \; \; \; ,
\nonumber \\
\hbar \delta \dot{\Pi}_{1,{\bf k}} &=& -\frac{1}{2} \left[ \hbar^{2}
k^{2}/2m_{1} + 2 \lambda_{1} n_{1} \right] \delta \rho_{1,{\bf k}}
- \lambda n_{1} \delta \rho_{2,{\bf k}} \; \; .
\label{e:dr}
\end{eqnarray}
Thus, the phase fluctuations of one condensate couple to the
density fluctuations of the other.
We cancel out the dependence on the phase fluctuations by taking the
derivative of the first equation in Eqs. (\ref{e:dr}) with respect to time
and by substituting $\delta \dot{\Pi}_{\bf k}$ from the second equation.
With $\delta \rho_{\bf k}(t) = \delta \rho_{\bf k} \cos(\Omega_{\bf k} t)$,  
we find the normal mode equations for the coupled density fluctuations,
\begin{eqnarray}
-\Omega^{2}_{\bf k} \delta \rho_{1,{\bf k}} &=&
- \omega_{1,{\bf k}}^{2} \delta \rho_{1,{\bf k}} - \lambda n_{1}
\frac{k^{2}}{m_{1}} \delta \rho_{2,{\bf k}} \; \; \; \; ,
\nonumber \\
-\Omega^{2}_{\bf k} \delta \rho_{2,{\bf k}} &=&
- \omega_{2,{\bf k}}^{2} \delta \rho_{2,{\bf k}} - \lambda n_{2}
\frac{k^{2}}{m_{2}} \delta \rho_{1,{\bf k}} \; \; \; \; ,
\label{e:nm}
\end{eqnarray}
where $\hbar \omega_{j,{\bf k}} =
\sqrt{ (\hbar^{2} k^{2}/2m_{j})^{2} + (\hbar^{2} k^{2}/m_{j})
n_{j} \lambda_{j}}$ (we assume $\lambda_{j} > 0$) denotes
the usual single condensate Bogoliubov dispersion. 
Requiring Eq.(\ref{e:nm}) to have non-trivial solutions
gives the dispersions of the double-condensate excitations:
\begin{eqnarray}
\Omega_{\pm, {\bf k}}^{2} &=& \frac{
\left[ \omega_{1,{\bf k}}^{2} + \omega_{2,{\bf k}}^{2} \right]}{2}
\nonumber \\
&& \; \; \; \pm \frac{
\sqrt{ \left[ \omega_{1,{\bf k}}^{2} - \omega_{2,{\bf k}}^{2} \right]^{2}
+ 4 (\lambda^{2}/
\lambda_{1} \lambda_{2}) c_{1}^{2} c_{2}^{2} k^{4} } } {2} \; \; \; .
\label{e:ndisp}
\end{eqnarray}
where $c_{j}$ is the sound velocity of the $j$-condensate, $c_{j}=\sqrt{n_{j}
\lambda_{j}/m_{j}}$.
The $\pm$-sign in the subscript, $\Omega_{\pm}$,
corresponds to the choice of $+$ or $-$ 
in Eq.(\ref{e:ndisp}).
The implications for the physics of the overlapping condensate
systems are profound:  the near-equilibrium
dynamics and thermodynamics
of the two-condensate system are essentially the dynamics and 
thermodynamics of two sets
of nearly non-interacting quasi-particles with energies 
$\Omega_{\pm {\bf k}}$.

        We are concerned with the stability of the
overlapping condensate system to which purpose we consider the long
wavelength limit, $k \rightarrow 0$, of the $\Omega$-dispersion relations.  
With the long wavelength Bogoliubov energies, $\omega_{j,{\bf q}}
\approx c_{j} q$, 
we find that the double condensate 
dispersions of Eq.(\ref{e:ndisp})
are also phonon-like,
$\Omega_{\pm,{\bf k}} \approx c_{\pm} k$ $(k \rightarrow 0)$, with
`sound velocities' $c_{+}$ and $c_{-}$ determined by
\begin{equation}
c_{\pm}^{2} = \frac{
\left[ c_{1}^{2} + c_{2}^{2} \right] \pm
\sqrt{ \left[  c_{1}^{2} - c_{2}^{2} \right]^{2} + 4 (\lambda^{2}/
\lambda_{1} \lambda_{2})
c_{1}^{2} c_{2}^{2}} }{2} \; \; \; \; .
\label{e:soundv}
\end{equation}
If $\lambda^{2} > \lambda_{1} \lambda_{2}$,
the inter-condensate interaction repels the energy levels so strongly
that $c_{-}^{2}$ becomes negative.  
Studying the normal mode frequencies 
of small deviations is a standard
test of stability and $c_{-}^{2} < 0$ indicates that the overlapping
condensate system is unstable against long wavelength excitations.

	The thermodynamic properties of the double condensate are determined
by minimizing the free energy $F$.  If the spatial variations of the
condensates are slow and the kinetic energy contributions may
be neglected, the zero-temperature free energy is the integral over
the free energy density $F({\bf r})$,
\begin{eqnarray}
F({\bf r}) &=& \frac{\lambda_{1}}{2} n_{1}^{2} ({\bf r}) +
 \frac{\lambda_{2}}{2} n_{2}^{2} ({\bf r}) +
\lambda n_{1}({\bf r}) n_{2} ({\bf r})
\nonumber \\
&& \;  \; \; \; \;
- \mu_{1}({\bf r)} n_{1}({\bf r})
- \mu_{2}({\bf r}) n_{2} ({\bf r}) \; \; \; ,
\label{e:f}
\end{eqnarray}
where the effective chemical potentials $\mu_{j} ({\bf r})$ include
the external potentials $v_{j} ({\bf r})$ experienced by the $j$ - bosons,
$\mu_{j}({\bf r}) = \mu_{j} - v_{j}({\bf r})$.  Minimizing F with
respect to the densities, $\delta F / \delta n_{j} ({\bf r}) = 0 $,
gives the Thomas-Fermi equations,
\begin{eqnarray}
\mu_{1}({\bf r}) &=& \lambda_{1} n_{1} ({\bf r}) + \lambda n_{2} ({\bf r})
\; \; \; ,
\nonumber \\
\mu_{2} ({\bf r}) &=& \lambda_{2} n_{2} ({\bf r}) + \lambda n_{1} ({\bf r})
\; \; \; .
\label{e:tf}
\end{eqnarray}
When $v_{j}({\bf r}) = 0$, the Thomas-Fermi condensate densities of
Eq.(\ref{e:tf}) are homogeneous.
However, equating first-order derivatives to zero,
only gives a minimum provided the second-order derivatives satisfy
$(\partial^{2} F / \partial n_{j}^{2}) > 0$ and $
(\partial^{2} F / \partial n_{1}^{2}) (\partial^{2} F / \partial n_{2}^{2})
- (\partial^{2} F / \partial n_{1} \partial n_{2} )^{2} > 0$.
The latter condition implies that the 
Thomas-Fermi equations (\ref{e:tf}) only gives
a minimum provided the stability criterion,
$\lambda^{2} < \lambda_{1} \lambda_{2}$, is 
satisfied.

	To see that `strongly' repulsive condensates, $\lambda 
> \sqrt{\lambda_{1} \lambda_{2}}$, lower their free 
energy by distributing the condensates inhomogeneously, we write 
the free energy density of Eq.(\ref{e:f})
in the absence of external potentials,
$v_{j} ({\bf r}) = 0$, as
\begin{eqnarray}
F({\bf r}) &=&
\frac{\lambda_{1}}{2} \left[ n_{1}({\bf r}) + n_{2}({\bf r})
\sqrt{\lambda_{2} / \lambda_{1} } \right]^{2} 
\nonumber \\
&& \; \; \; \;
+ \left[ \lambda - \sqrt{\lambda_{1} \lambda_{2}} \right]
n_{1} ({\bf r}) n_{2} ({\bf r})
\nonumber \\
&& \; \;  \; - \mu_{1} n_{1} ({\bf r})
- \mu_{2} n_{2} ({\bf r})  \; \; \; \; .
\label{e:f2}
\end{eqnarray}
Starting from the homogeneous overlapping condensate system, 
redistributing bosons 1 and 2 spatially 
while keeping
$[ n_{1} ({\bf r}) + n_{2} ({\bf r}) \sqrt{\lambda_{2} / \lambda_{1} } ]$
constant over space,
can lower the energy by decreasing the overlap
integral $\int d^{3} r \; n_{1} ({\bf r}) n_{2} ({\bf r})$.  The lowest value
is reached by spatially separating the two condensates.
In that case, although the hamiltonian is translationally invariant
($v_{j}({\bf r}) = 0$), the double condensate is not: the double
condensate system spontaneously breaks translational symmetry.
Note that the phase separation is consistent with the above discussed
dynamical instability -- the separated condensate system has no region
that can be described locally as two homogeneous condensates.  In contrast,
strongly attractive condensates 
($\lambda < 0, \lambda^{2} > \lambda_{1} \lambda_{2}$)
decrease the free energy by increasing their mutual overlap
and overlapping condensates are unstable.  
This is similar to the behavior of single condensates of negative
scattering length, and we expect the strongly attractively double 
condensates to be similarly unstable.
The same analogy suggests that a confining potential might give stable or 
metastable strongly attractive double condensates.

	The free energy of the phase separated condensate system in 
a macroscopic volume $V$, is the sum of the single condensate
free energies, 
condensate 1 confined to a volume $V_{1}$, and condensate 2
confined to $V-V_{1}$.  Minimizing the total free energy with respect to
$V_{1}$ gives the equilibrium condition of equal pressures exerted by
both condensates.  With the pressure $P_{j} = \lambda_{j}
n_{j,s}^{2}/2$, where $n_{j,s}$ denotes the density within the separated
condensates, we are lead to the equivalent condition for the condensate
densities $n_{1,s} =
n_{2,s} \sqrt{\lambda_{2}/\lambda_{1}} $. 

        We note that the assumption of slowly varying condensate wavefunctions,
necessary to justify neglecting the kinetic energy in the free energy of
Eq.(\ref{e:f}), is violated at the boundary of the two condensates.  In fact,
an infinitely sharp boundary gives an infinite kinetic energy contribution.
The effect of the kinetic energy is then to give a boundary region
of finite size b, in which the wavefunctions smoothly tend to zero
as the condensates cross the boundary.  
We assume that the condensates are so large that the boundary region
which separates them can be approximated locally as a planar region with
densities that vary spatially as functions of the coordinate $z$
with the $z$-axis perpendicular to the boundary surface of area $A$.
The kinetic energy contribution,
$E_{kin}(b)$, is then approximately equal to
$E_{kin}(b) \approx (A \hbar^{2} n_{1,s}/2 m_{1} b)
[ 1+ \sqrt{\lambda_{1}/\lambda_{2}} (m_{1}/m_{2}) ]$,
where we used that $(n_{2,s}/n_{1,s}) = 
\sqrt{\lambda_{1}/\lambda_{2}}$.  The
overlap of the condensates in the boundary region increases the interaction
energy by an amount $E_{int}(b)$, which we estimate by modeling the
condensate densities in the boundary region, $z \in (0,b)$, crudely as
$n_{1}(z) \approx n_{1,s} (b - z )/b$ and $ n_{2}(z) 
\approx n_{2,s} z/b$.  With
Eq.(\ref{e:f2}), we find that $E_{int}(b) \approx A
(\lambda - \sqrt{\lambda_{1} \lambda_{2}}) n_{1,s} n_{2,s} b /6$.  
To estimate the actual boundary size $\overline{b}$,
we minimize the boundary energy, $E_{b}(b) = E_{kin}(b) + 
E_{int}(b)$ with
respect to $b$ and find
\begin{equation}
\overline{b} =
2 l_{1} \sqrt{3}  
\sqrt{ 
\frac{ [ 1 + (m_{1}/m_{2}) \sqrt{\lambda_{1}/\lambda_{2}} ]}
{ \left[ \lambda / \sqrt{\lambda_{1} \lambda_{2}} -1 \right] } 
} \; \; ,
\label{e:b}
\end{equation}
where $l_{1}$
is the coherence length of condensate 1,
$l_{1} = \hbar/\sqrt{4 m_{1} n_{1,s} \lambda_{1}}$.  The boundary contribution
to the energy is a surface energy
$E_{b} (\overline{b}) = \sigma A$, where the surface tension 
$\sigma$ is proportional to the coherence length $l_{1}$, 
to the pressure $P_{1}$,
$P_{1} = n_{1,s}^{2} \lambda_{1}/2$, and to a dimensionless constant
$\Sigma_{1}$ which depends solely on the mass and  
interaction strength ratios, 
$\Sigma_{1} = 4 \sqrt{ [ 1 + (m_{1}/m_{2}) \sqrt{\lambda_{1}/\lambda_{2}} ]
[ \lambda / \sqrt{\lambda_{1} \lambda_{2}} -1 ] }/\sqrt{3}$,
$\sigma = l_{1} P_{1} \Sigma_{1}$.

        In the absence of external potentials, a `droplet' of condensate 1
immersed in a much larger condensate 2 of density $n_{2}$ 
minimizes 
$E_{b}$ by taking on the shape of a sphere of radius $R$, $V_{1}
= (4\pi/3) R^{3}$.  
We can now imagine creating the double condensate system starting from
a single condensate of type 2 and replacing condensate 2 bosons 
in the droplet volume $V_{1}$ by
condensate 1 bosons.  The energy $\Delta E$ required in the replacement
is equal to $\Delta E = [ \lambda_{1,s} n_{1}^{2}/2 -
\lambda_{2} n_{2,s}^{2}/2 ] V_{1} + E_{b}$.  Minimizing the
`replacement energy' $\Delta E$
with respect to $V_{1}$ and realizing that $E_{b} \propto V_{1}^{2/3}$,
we find 
\begin{equation}
\frac{\lambda_{1} n_{1,s}^{2}}{2} =
\frac{\lambda_{2} n_{2,s}^{2}}{2} + \frac{2}{3} \;
\frac{E_{b}}{V_{1}} \; .
\label{e:nn}
\end{equation}
The previous result, $\lambda_{1} n_{1,s}^{2}/2 =
\lambda_{2} n_{2,s}^{2} /2$, obtained by
ignoring the boundary energy, is accurate provided the size
of the droplet exceeds $R_{s} = 2 \sigma / P_{1}$ = $ 2 l_{1} \Sigma_{1}$.
The energy per droplet particle, $\Delta E/N_{1}$, with Eq.(\ref{e:nn})
is equal to
$\Delta E/N_{1} = 5 E_{b}/V_{1} = 5 \sigma/[R n_{1,s}]$, 
a function that decreases monotonically as $N_{1}$ increases.
Consequently, splitting up the droplet into smaller
droplets further increases the free energy and it is
energetically favorable for
condensate 1 to gather in a single region of space
(i.e. real space condensation).

	To describe separated
double condensates in traps, we subtract the overlap term, 
$\lambda n_{1} n_{2}$, and include
the boundary surface energy in the expression of the free energy
(Eq.(\ref{e:f})).  
The validity of this description
rests on two conditions: 1. the local coherence length
within each condensate is much less than the length scale on which
the condensates vary spatially and 2.  the change of the potential energy
across the inter-condensate boundary, $|{\bf f}_{j}| \overline{b}$, where
${\bf f}_{j} = - \nabla v_{j}$ represents the external 
force experienced by bosons j near the boundary,
is much less than the local chemical potential
$|{\bf f}_{j}| \overline{b} << \lambda_{j} n_{j,s}$ 
($j=1,2$). If these conditions are satisfied, 
the physics of the phase separation is similar to the above $v_{j} ({\bf r}) = 
0$-case, and we can answer interesting
questions regarding trapped phase separated condensates.  For instance:
if we add a droplet of condensate 1 to a trapped
condensate 2, does it `sink' to the middle of the trap,
or does it remain `floating' on the surface of condensate 2?
For the sake of simplicity, we assume that the size of the droplet
is large enough to neglect the boundary surface energy and small
enough to neglect the spatial variation of the density inside
the droplet.  Then, the previously defined `replacement energy' 
$\Delta E$ depends on the
center of mass position ${\bf R}$ of the droplet through the
external potentials, $v_{1}({\bf r}) = v({\bf r})$, and $v_{2}({\bf r})
= \alpha v({\bf r})$.
Since the pressures
inside and outside the droplet are equal, $\lambda_{1} n_{1,2}^{2}/2
- \lambda_{2} n_{2,s}^{2}/2 \approx 0$, we find that 
\begin{eqnarray}
\Delta E ({\bf R}) &=& \int_{ V_{1} }
[ n_{1} ({\bf r}) v_{1}({\bf r}) - n_{2} ({\bf r}) v_{2} ({\bf r}) ]
\; d^{3} r
\nonumber \\
&\approx& N_{1} v ({\bf R}) [ 1 - \alpha 
\sqrt{\lambda_{1}/\lambda_{2}} ] \; \; \; .
\label{e:potdrop}
\end{eqnarray}
Thus, even though both bosons experience a trapping potential, 
if $\alpha \sqrt{\lambda_{1}/\lambda_{2}} > 1$, the force on the droplet,
$- N_{1} [1-\alpha \sqrt{\lambda_{1}/
\lambda_{2}}] \nabla v$, is directed outwards : the
droplet `floats'.

	The thinner the layer of the floating condensate, the more
important are the effects of the boundary surface energy.  Indeed, with a 
few condensate 1 particles, the energy will be minimized by
covering only part of the surface of condensate 2 (thereby reducing
the surface boundary energy).  Here, we only consider the case where 
enough bosons have been added for condensate 1 to 
`wrap' around condenate 2, and 
we can ignore boundary surface
energy effects.  Within the single condensate regions, the condensates
are described in the Thomas-Fermi approximation of Eq.(\ref{e:tf}) (putting
$\lambda = 0$), $n_{1}({\bf r}) = [\mu_{1}-v({\bf r})]/\lambda_{1}$ and
$n_{2}({\bf r}) = \alpha [\mu_{2} - v({\bf r})]/\lambda_{2}$, where
$\mu_{1}$ and $\alpha \mu_{2}$ are the chemical potentials.  
In that case, the boundary surface is the equipotential
surface $v({\bf R}) = \mu_{b}$ where the pressures of both condensates,
$\lambda_{j} n_{j}^{2}({\bf r})/2$, are equal.
This leads to 
\begin{equation}
\mu_{b} = \mu_{2} - \frac{\left[ \mu_{1} - \mu_{2} \right]}{ \left[ \alpha
\sqrt{\lambda_{1} / \lambda_{2} } -1 \right] } \; \; \; .  
\label{e:mb}
\end{equation}
In figure 1, we show
a typical density profile for two separated
condensates in a spherically symmetric
trap.
With $\lambda_{j} n_{j}(R_{b}) = \mu_{j} - v_{j}(R_{b})$, we find for
the case shown in Fig.(1) that $\mu_{1} - \mu_{2} = \lambda_{1} n_{1}(R_{b})
- \lambda_{2} n_{2}(R_{b})/\alpha > 0$. Using 
$[n_{1}(R_{b})/n_{2}(R_{b})] = \sqrt{\lambda_{2}/\lambda_{1}}$ we then find 
that $\alpha \sqrt{\lambda_{1}/\lambda_{2}} > 1$, so that condensate 1
should indeed float on top of condensate 2.
Of course, the experimentally relevant quantities are the number
of boson particles, $N_{1}$ and $N_{2}$, rather than $\mu_{1}$ and $\mu_{2}$.
The chemical potentials can be determined by inverting
$N_{1} = \int_{V_{1}} d^{3} r \;
n_{1} ({\bf r})$ and $N_{2} = \int_{V_{2}} d^{3} r \; n_{2}({\bf r})$,
where the boundary between $V_{1}$ and $V_{2}$ is defined by 
Eq.(\ref{e:mb}), to give $N_{j}(\mu_{1},\mu_{2})$.

	In summary, we have shown that the homogeneous overlapping
double condensate system with strong inter-condensate interactions
($\lambda^{2} > \lambda_{1} \lambda_{2}$) is unstable.
In that case, attractive condensates ($\lambda < 0$) collapse and
repelling condensates ($\lambda > 0$) separate spatially.
In between the separated condensates is a region of partial overlap.
We have estimated the size of this region, as well as the resulting surface
tension.  Finally, we briefly discussed phase separation of large condensates
in atomic traps.

	The phase separation suggests many experimental applications.
For instance, one can use a condensate to spatially confine a droplet
of a different condensate in the middle of a trap.  Using light
that is resonant with the droplet atoms, the droplet can be displaced
and its subsequent motion inside the confining condensate can be observed.  
If the motion is undamped, we have a direct observation of superfluidity.
To further motivate such experiments, we mention that damping 
generally could occur at 
velocities less than the sound velocity of the confining condensate
because of the creation of superfluid vortices etc...
Furthermore, it is interesting
to note that schemes have been proposed to continuously alter the
interaction strengths, either by using light \cite{Kagan}, or by
varying the bias field of the magnetic traps \cite{Verhaar}, 
so that it might be
possible in the future to observe phase separation in real time as 
the interaction strengths are altered.

	The author gratefully acknowledges fruitful interactions
with Dr. P. Tommasini, Prof. E. Heller, Prof. A. Dalgarno and
Prof. K. Huang.
The work of the author is supported by the NSF through a grant for the
Institute for Atomic and Molecular Physics at Harvard University
and Smithsonian Astrophysical Observatory.

\newpage
\centerline{\bf Figure Captions}
\noindent
\underline{Fig.1 }:
	Plot of a typical phase separated double condensate in
a spherically symmetric trap.  
In reality, the boundary of condensates 1 and 2 (at $R = R_{b}$ where $v(R_{b})
= \mu_{b}$) is not infinitely sharp and the condensates overlap over
a region of size $b$ that is estimated in the text.



\begin{thebibliography}{99}
\bibitem{rem} Previously, the only candidate for creating interpenetrating
bosonic superfluids seemed to be a mixture of $^{4}$He and $^{6}$He.
Unfortunately, the short lifetime of the $^{6}$He-isotope ($\sim 0.8$ seconds)
did not allow cooling to sufficiently low temperatures. Nevertheless,
the prospect motivated theoretical work that is relevant to our
paper \cite{khal},
\cite{bas}, \cite{nep} and \cite{fet}.
\bibitem{remark} The same equations can be obtained from the Heisenberg
equations of motion, neglecting the interactions between excited particles.
\end{thebibliography}
\end{document}